# Minimizing CGYRO HPC Communication Costs in Ensembles with XGYRO by Sharing the Collisional Constant Tensor Structure

Igor Sfiligoi
University of California San Diego
La Jolla, CA, USA
isfiligoi@sdsc.edu

Emily A. Belli
General Atomics
La Jolla, CA, USA
bellie@fusion.gat.com

Jeff Candy
General Atomics
La Jolla, CA, USA
candy@fusion.gat.com

## ABSTRACT

First-principles fusion plasma simulations are both compute and memory intensive, and CGYRO is no exception. The use of many HPC nodes to fit the problem in the available memory thus results in significant communication overhead, which is hard to avoid for any single simulation. That said, most fusion studies are composed of ensembles of simulations, so we developed a new tool, named XGYRO, that executes a whole ensemble of CGYRO simulations as a single HPC job. By treating the ensemble as a unit, XGYRO can alter the global buffer distribution logic and apply optimizations that are not feasible on any single simulation, but only on the ensemble as a whole. The main saving comes from the sharing of the collisional constant tensor structure, since its values are typically identical between parameter-sweep simulations. This data structure dominates the memory consumption of CGYRO simulations, so distributing it among the whole ensemble results in drastic memory savings for each simulation, which in turn results in overall lower communication overhead.

## 1 Introduction

CGYRO [1] is a popular first-principles fusion turbulence simulation tool, used to validate basic fusion theory, plan experiments, interpret results on existing fusion devices, and ultimately to design future devices. Simulations are both compute and memory intensive, but it is the memory requirements that directly drive the need for multi-node HPC on modern hardware. While CGYRO can linearly scale compute over multiple nodes, communication overheads do increase with node count [2].

CGYRO implements the complete Sugama electromagnetic gyrokinetic theory [3], which results in a very large collisional constant tensor structure (*cmat*), which is computed once per simulation and then used for computing the simulation's collisional step [4]. This approach, which trades memory intensity for lower compute cost, does drastically increase the memory usage but allows for order of magnitude compute speedup in the collision step, which uses an implicit time-stepping algorithm. To illustrate the magnitude of memory used, for the benchmark input *nl03c* the constant *cmat* is 10x the size of all the other memory buffers combined.

While there is very little that can be done to reduce the memory needs of a single CGYRO simulation, combining several simulations into a single ensemble HPC job does open new opportunities. A careful analysis of *cmat* construction shows that only a subset of the input parameters influences its value, and there are many fusion studies that do not change them between simulation runs. This realization inspired the development of the XGYRO tool, which leverages the CGYRO codebase, but allows for the execution of multiple fusion simulations as an ensemble which shares a single copy of the constant *cmat*. This obviously drastically reduces the per-simulation memory needs, thus allowing for running most of the compute of each simulation on fewer HPC nodes, which in turn results in higher computational efficiency. The net result is more simulations completed on the same compute budget.

## 2 Fusion simulation problem splitting

CGYRO fusion simulations mostly operate on 3D tensor buffers, with the three dimensions labelled as *configuration* (*nc*), *velocity* (*nv*) and *toroidal* (*nt*). The constant *cmat* tensor structure is an exception, in that it requires a 4D tensor of size (*nv* x *nv* x *nc* x *nt*), which explains why it is so much bigger than all the other buffers. Data partitioning happens by splitting and distributing the tensors in all but one dimension. There are three logical phases during simulation, conventionally called streaming (*str*), non-linear (*nl*) and collisional (*coll*), with *str* requiring the complete *nc* dimension, *nl* requiring the complete *nt* dimension, and *coll* requiring the complete *nv* dimension. MPI collective communications are used to transition between the various partitioning schemes. For the purpose of this document, we will mostly ignore the *nl* phase, as there is never a direct transition from it to the *coll* phase.

On a more technical side, the MPI communicator that splits the *nv* dimension in the *str* phase is used in CGYRO for two distinct purposes. It is indeed used to perform a transpose of the buffers between the *str* and *coll* phases, using the MPI `AllToAll` collective communication libraries. But it is also used several times for aggregating results from partial transforms between processes during the *str* phase itself (also known as *field* and *upwind*), using the MPI `AllReduce` collective libraries, as shown in Figure 1. This reuse is possible since it involves the same processes in both cases.

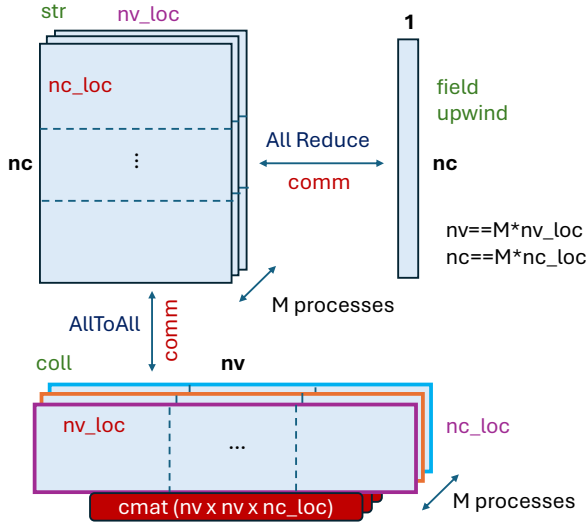

**Figure 1: CGYRO *str* and *coll* communication logic.**

As for the *cmat* constant tensor structure, it is only used in the *coll* phase. It is thus split and distributed among the processes in the same way the *nc* dimension is split and distributed. The size of cmat in each process is thus (*nv* x *nv* x *nc_loc*), for each *toroidal* slice. The relative difference in size compared to the other buffers thus does not change with strong scaling, i.e. when nc_loc becomes smaller.

## 2.1 Ensemble simulations with XGYRO

The XGYRO tool was developed to run several independent simulations as an ensemble, wrapped up as a single HPC job. To minimize new code development, it is implemented as a thin MPI initialization and partitioning layer around the CGYRO codebase, with minor changes to the latter.

XGYRO uses a different data partitioning logic for the constant *cmat*, compared to the other tensor buffers. The former is distributed across all the processes in the ensemble, while all the others use the standard per-simulation distribution logic. This required the separation of the MPI communicator that handles the *nv* splitting in *str* from the one handling *coll*, as the number of processes involved differs between the two, as shown in Figure 3. Most of the other code remained unchanged.

Since *cmat* is now shared between all the simulations in an ensemble, its size does not change if we change the number of simulations in a XGYRO ensemble. And since all other buffers do grow linearly with the number of simulations, *cmat*'s relative memory consumption proportionally decreases. This means that we can use significantly fewer HPC nodes to run the same number of simulations in parallel.

Looking at communication patterns, we observe that the major change, as we vary the ensemble size, is the number of processes participating in the MPI `AllReduce` collective communication. Since the overall cost of `AllReduce` is proportional with the number of participating processes, this should result in significant reduction in communication cost as we increase the number of simulations per ensemble.

## 3 XGYRO benchmark results

To validate the performance benefits of XGYRO, we compare the runtime of running 8 variants of the *nl03c* benchmark simulations using 32 OLCF Frontier HPC nodes, either sequentially with CGYRO or as an ensemble with XGYRO. Note that a single CGYRO simulation does require at least 32 nodes.

The benchmark data shown in Figure 2 validate our assumptions, showing that, for a simulation reporting step, XGYRO required 250s while the sum of 8 independent CGYRO simulations took 375s. This represents a 1.5x speedup. The major difference, as expected, is the time spent performing the *str* communication, 33s for XGYRO vs 145s for the CGYRO sum. Complete simulation logs can be found in [5].

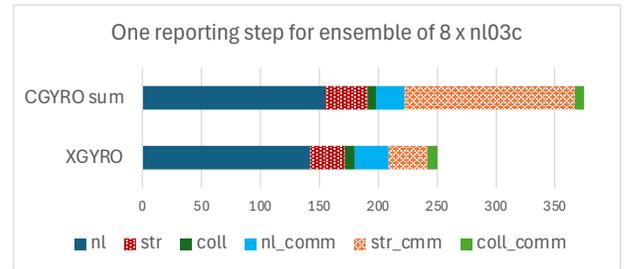

**Figure 2: Benchmark results for CGYRO and XGYRO at t=81. All numbers in seconds per reporting time step.**

## ACKNOWLEDGMENTS

This material is based upon work supported by the U.S. Department of Energy, Office of Science, Office of Fusion Energy Science under awards DE-FG02-95ER54309 and DE-SC0024425 (FRONTIERS SciDAC-5 project). An award of computer time was provided by the INCITE program. This research used resources of the Oak Ridge Leadership Computing Facility, which is an Office of Science User Facility supported under Contract DE-AC05-00OR22725. Computing resources were also provided by the National Energy Research Scientific Computing Center, which is an Office of Science User Facility supported under Contract DE-AC02-05CH11231.

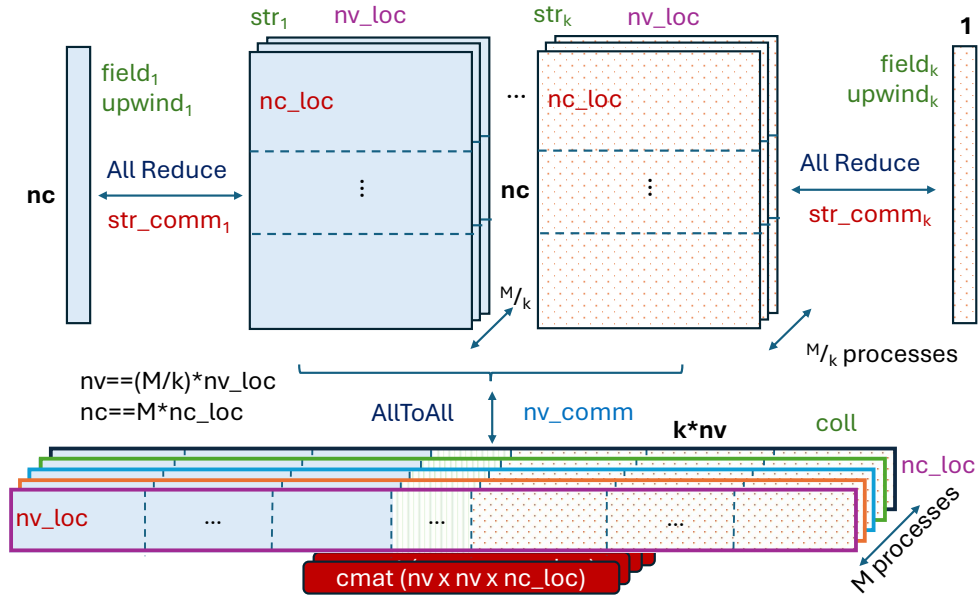

**Figure 3: XGYRO *str* and *coll* communication logic of an ensemble of k CGYRO simulations sharing *cmat*.**